\documentclass[aps,pra,floatfix,superscriptaddress,preprint, showpacs]{revtex4}

\usepackage{graphicx}
\usepackage{graphics}
\usepackage{amssymb}
\usepackage{amsmath}
\usepackage{epsfig}
\usepackage{latexsym}
\usepackage{color}
\usepackage{rotating}
\usepackage{subfigure}

\begin{document}

\title{Quantum noise in a nano mechanical Duffing resonator.}

\author{E. Babourina-Brooks, A. Doherty, G. J. Milburn}
\affiliation{Department of Physics, School of Physical Sciences, The University of Queensland, St Lucia, QLD 4072, Australia}

\begin{abstract}
We determine the  small signal gain and noise response of an amplifier based on the nonlinear response of a quantum nanomechanical resonator. The resonator is biased in the nonlinear regime by a strong harmonic bias  force and we determine the response to a small additional driving signal detuned with respect to the bias force. 
\end{abstract}
\pacs{85.85.+j,07.10.Cm,42.50.Lc}

\maketitle

\section{Introduction}
Almog et al.\cite{Buks} recently demonstrated classical noise squeezing and amplification in a nanomechanical resonator with a significant fourth order nonlinearity in the elastic  potential energy. This system is dynamically equivalent to the Duffing oscillator, which exhibits a fixed point bifurcation: a loss of stability in a fixed point as a control parameter (in this case a driving force)  is varied\cite{Kozinsky}.  Near the bifurcation, the system becomes very sensitive to fluctuations, a sensitivity that can be harnessed to make a signal amplifier\cite{Wisenenfeld}, or sensitive charge detector\cite{Kromer}. Such a bifurcation amplifier has recently been used in the study of superconducting qubits\cite{Siddiqi}. In this paper we calculate the small signal gain and noise power spectrum for quantised quartic nonlinear nanomechanical oscillator when the harmonic driving force is subject to a perturbation.  Our treatment complements the study of Buks and Yurke\cite{BuksYurke} in which the response of a nonlinear resonator to changes in the linear frequency, due to a mass perturbation,   enable a sensitive mass detection transducer.

The Hamiltonian for the nanomechanical system is\cite{Carr}
\begin{equation}
H=\frac{p^2}{2m^*}+\frac{m^*\omega_0^2}{2}(x^2+\frac{\alpha}{2} x^4)
\end{equation}
where $m^*$ is the effective mass of the nanomechanical resonator, $\omega_0$ is the linear resonator frequency taking into account the applied strain. Kozinsky et al.\cite{Kozinsky} give the nonlinear parameter as
\begin{equation}
\alpha=\frac{2\sqrt{3}}{9a_c^2 Q}
\end{equation}
where $Q$ is the quality factor of the resonator and $a_c$ is the critical amplitude at which the resonance amplitude has an infinite slope as a function of the driving frequency.



The position and momentum operators of  the nanomechanical resonator may be written in terms of raising and lowering operators, $a^\dagger, a$,
\begin{eqnarray}
x & = & \sqrt{\frac{\hbar}{2m\omega_0}}(a+a^\dagger)\\
p & = &  -i\sqrt{2\hbar m\omega_0}(a-a^\dagger)
\end{eqnarray}
We also include a harmonic driving field at frequency $\omega_p$, and amplitude $\epsilon_p$, which  we call the pump, that is used to set the operating conditions of the device, and a weaker harmonic signal driving field at frequency $\omega_s$ and amplitude $\epsilon_s$.  The Hamiltonian may then be written as 
\begin{equation}
H=\hbar\omega_0a^\dagger a+\hbar\frac{\chi}{6}(a+a^\dagger)^4+\hbar(2\epsilon_p\cos(\omega_p t)+2\epsilon_s\cos(\omega_s t))(a+a^\dagger)
\end{equation}
where $\chi$ gives the nonlinear dispersion. It is given in terms of the nonlinearity parameter $\alpha$ as
\begin{equation}
\chi=\frac{3\hbar\alpha}{8m_*}
\end{equation}
In the example of a doubly clamped platinum beam in \cite{Kozinsky} , $\chi\sim 3.4\times 10^{-4}\ \mbox{s}^{-1}$.

Moving to an interaction picture at the pump frequency, $\omega_p$, and assuming that $\omega_0, \omega_p>>\chi$ to neglect rapidly oscillating terms in the quartic term, the Hamiltonian becomes 
\begin{equation}
H_I=\hbar\Delta a^\dagger a+\hbar\chi(a^\dagger)^2a^2+\hbar\epsilon_p(a+a^\dagger)+\hbar\epsilon_s(a e^{i\delta t}+a^\dagger e^{-i\delta t})
\end{equation}
where $\Delta=\omega_0-\omega_p$, $\delta=\omega_s-\omega_p$ are the detuning of the resonator and the signal from the pump respectively. 

 The actual physical model for dissipation in quantum NEMS is still under investigation. For the purposes of this paper we will model damping via the quantum optics master equation\cite{Walls-GJM}. This assumes that the nanomechanical resonator is under-damped due to a weak coupling to a bath of harmonic oscillators. It further assumes the validity of the rotating wave approximation for the interaction between the NEMS degree of freedom and the bath oscillators which is expected to be a good approximation for sufficiently high NEMS frequency and sufficiently weak coupling to the bath. 
 
The master equation in the interaction picture for the system is then given by
\begin{equation}
\frac{d\rho}{dt}= -\frac{i}{\hbar}[H_I,\rho]+\frac{\gamma}{2}(\bar{n}+1)(2a\rho a^\dagger -a^\dagger a \rho-\rho a^\dagger a)+\frac{\gamma\bar{n}}{2}(2a^\dagger \rho a- aa^\dagger\rho-\rho aa^\dagger)
\end{equation}
where $\gamma$ is the rate of energy loss from the resonator and $\bar{n}$ is the mean phonon number in a bath oscillator at frequency $\omega_0$. We will henceforth assume low temperature operation so that $\bar{n}\approx 0$.   In the case of no signal field, this model was introduced long ago in quantum optics to describe optical bistability due to a Kerr nonlinear medium\cite{DrumWalls}. The system has a steady state, or fixed point, which can change stability as the driving field is varied. It is this dependance of fixed point stability on driving field that can be used to amplify a weak driving signal.

In addition to this dissipative channel,  the transducer for the nanomechanical displacement  itself provides an open channel by which the resonator is coupled to the external world. A variety of transducers are possible, including single electron transistors\cite{Schwab}, super-conducting single electron transistors\cite{Armour} and super-conducting co-planar microwave cavities\cite{VitWoll}. We do not model a specific device here but proceed with a more generic description. 

We will assume that the transducer is described by a multi mode bosonic field,  with positive frequency components, $a_{in}(t), a_{out}(t)$,  and a carrier frequency close to the pump field frequency of the resonator. If we further assume that this coupling is linear and make the rotating wave and Markov approximation, the bosonic field after the interaction, $a_{out}(t)$, with the resonator is related to the complex amplitude of the resonator and the input field to the transducer $a_{in}(t)$ by\cite{Walls-GJM}
\begin{equation}
a_{out}(t)=\sqrt{\gamma_T}a(t)-a_{in}(t)
\label{input-output}
\end{equation}
where $a_{in}(t)$ is the quantum noise operator for the input noise to the transducer field, which for simplicity we will take to be in a coherent state with a monochromatic component at carrier frequency $\omega_s$ and amplitude $\epsilon_s$. This assumes that the carrier frequency of the transducer field is large compared to $k_BT/\hbar$.  We have in mind a transducer in the form of a microwave field in a superconducting co-planar strip line \cite{Lehnert}. At a carrier frequencies of GHz and milliKelvin temperatures  such a transducer may be regarded as in a coherent state. Under these assumptions we see that the transducer field is simply proportional to the operator $a(t)$ for the nanomechanical resonator. The total damping rate, $\gamma$ should be regarded as including the transducer damping rate $\gamma_T$.
 
 In order to see the displacement of the resonator we need to make a phase sensitive measurement of the transducer field. This requires a frequency and phase reference. In the case of electromagnetic fields this would be done using homodyne detection of a quadrature phase amplitude of the field.   We are thus led to define the {\em quadrature phase amplitude} operators for a nanomechanical resonator in the Heisenberg picture as
\begin{equation}
\hat{X}_\theta(t)\equiv a(t) e^{i(\theta+\omega_{lo}t) }+ a^\dagger(t) e^{-i(\theta+\omega_{lo}t) }
\end{equation}
where $\theta$ and $\omega_{lo}$ are the phase and frequency reference provided by some form of local oscillator.  In our case, the input signal to the amplifier is set to $\omega_s$. As we will see, this leads to a long time solution in which the signal has Fourier components at frequencies $\omega_p\pm\delta$.   With this in mind we then assume that the local oscillator frequency is set so that $\omega_{LO}=\omega_p\pm\delta$.    The operators $a(t), a^\dagger(t)$ appearing in these expressions are multi-mode operators describing the quantum noise of the detected signal propagating away from the local system. It is conventional to write them in terms of the positive frequency components $a(\omega)$ as
\begin{equation}
a(t)=\int_0^\infty d\omega a(\omega) e^{-i\omega t}
\end{equation}
where $
[a(\omega),a^\dagger(\omega')]=\delta(\omega-\omega')$. The quadrature operators $\hat{X}_\theta(t)$ and $\hat{X}_{\theta+\pi/2}(t)$ are canonically conjugate multimode operators with equal-time commutation relations,
\begin{equation}
[ \hat{X}_\theta(t),\hat{X}_{\theta+\pi/2}(t)]=2i
\end{equation}
We are working in an interaction picture defined with respect to the pump frequency $\omega_p$ so we can write 
 \begin{equation}
\hat{X}_\theta(t)\equiv a_I(t) e^{i(\theta\pm\delta t) }+ a_I^\dagger(t) e^{-i(\theta\pm\delta t) }
\end{equation}
where $a_I(t)=a(t)e^{i\omega_p t}$ and we have defined the local oscillator frequency as $\omega_{LO}=\omega_{p}\pm\delta$.  In terms of the positive frequency components the interaction picture amplitude may be written as 
\begin{equation}
a_I(t)=\int_0^\infty d\omega a(\omega) e^{-i(\omega-\omega_p)t}
\end{equation}
We expect that the states of interest are such that $n(\omega)=\langle a^\dagger(\omega) a(\omega)\rangle$ is significantly different from zero only over a bandwidth $B$ centered on $\omega=\omega_p$, with $\omega_p >>B$. In an experiment $B$ is the order of the resonator line width which is of order MHz,  and $\omega_p$ is at GHz. We thus make a change of variable $\epsilon=\omega-\omega_p$, and setting $\omega_p\rightarrow -\infty$ we can write 
\begin{equation}
a_I(t)=\int_{-\infty}^\infty d\epsilon \tilde{a}(\epsilon)e^{-i\epsilon t}
\end{equation}
indicating that $ \tilde{a}(\epsilon)\equiv a(\omega_p+\epsilon)$ is the Fourier transform of the interaction picture operator $a_I(t)$. In a similar way we find that 
\begin{equation}
a_I^\dagger (t)=\int_{-\infty}^\infty d\epsilon \tilde{a}^\dagger(\epsilon)e^{-i\epsilon t}
\end{equation}
but with the definition $ \tilde{a}^\dagger(\epsilon)\equiv a(\omega_p-\epsilon)$. Note that  the commutation relations for the Fourier components is not the same as that for the original positive and negative frequency components,  
\begin{equation}
 [\tilde{a}(\epsilon),\tilde{a}^\dagger(\epsilon')]=\delta(\epsilon+\epsilon')
 \end{equation}

\section{Bistability}
We first establish the operating conditions for the device when there is no signal present, $\epsilon_s=0$. Under certain conditions, the energy in the nano mechanical resonator  as a function of the driving intensity can exhibit multiple stable fixed points and hysteresis. We will follow the approach of Drummond and Walls\cite{DrumWalls} which is based on the positive P representation of the density operator. In the semiclassical approximation, the equation of motion for the mean amplitude, $\alpha\equiv \langle a\rangle$, is given by
\begin{equation}
 \dot{\alpha} = -i\epsilon_p-\left (\gamma/2+i(\Delta+2\chi|\alpha|^2)\right )\alpha
 \label{semiclassical}
 \end{equation}
 where the dot indicates a time derivative. 
 The fixed point (or semiclassical steady state) is defined by $\dot{\alpha}=0$, which corresponds to a complex amplitude $\alpha_0$ must satisfy
 \begin{equation}
I_p=n_0\left [\frac{\gamma^2}{4}+(\Delta+2\chi n_0)^2\right ]
 \label{state-eqn}
 \end{equation}
where $I_p= \epsilon_p^2$ is proportional to the pump power driving the nanomechanical resonator and  $n_0=|\alpha_0|^2$ determines the average energy in the nanomechanical resonator by $E=\hbar\omega_0 n$. Considered as a function of $n_0$, $I_p$ is a cubic with turning points at the values of $n_0$ that satisfy 
\begin{equation}
\frac{dI_p}{dn_0}=\frac{\gamma^2}{4}+(\Delta+6\chi n_0)(\Delta+2\chi n_0)=0
\label{turning}
\end{equation}
However when we regard $n_0$ as a function of the pump power, it is multi-valued and Eq.(\ref{turning}) defines values at which the slope diverges, indicative of a change in stability.

\begin{figure}[h]
\includegraphics[scale=0.5]{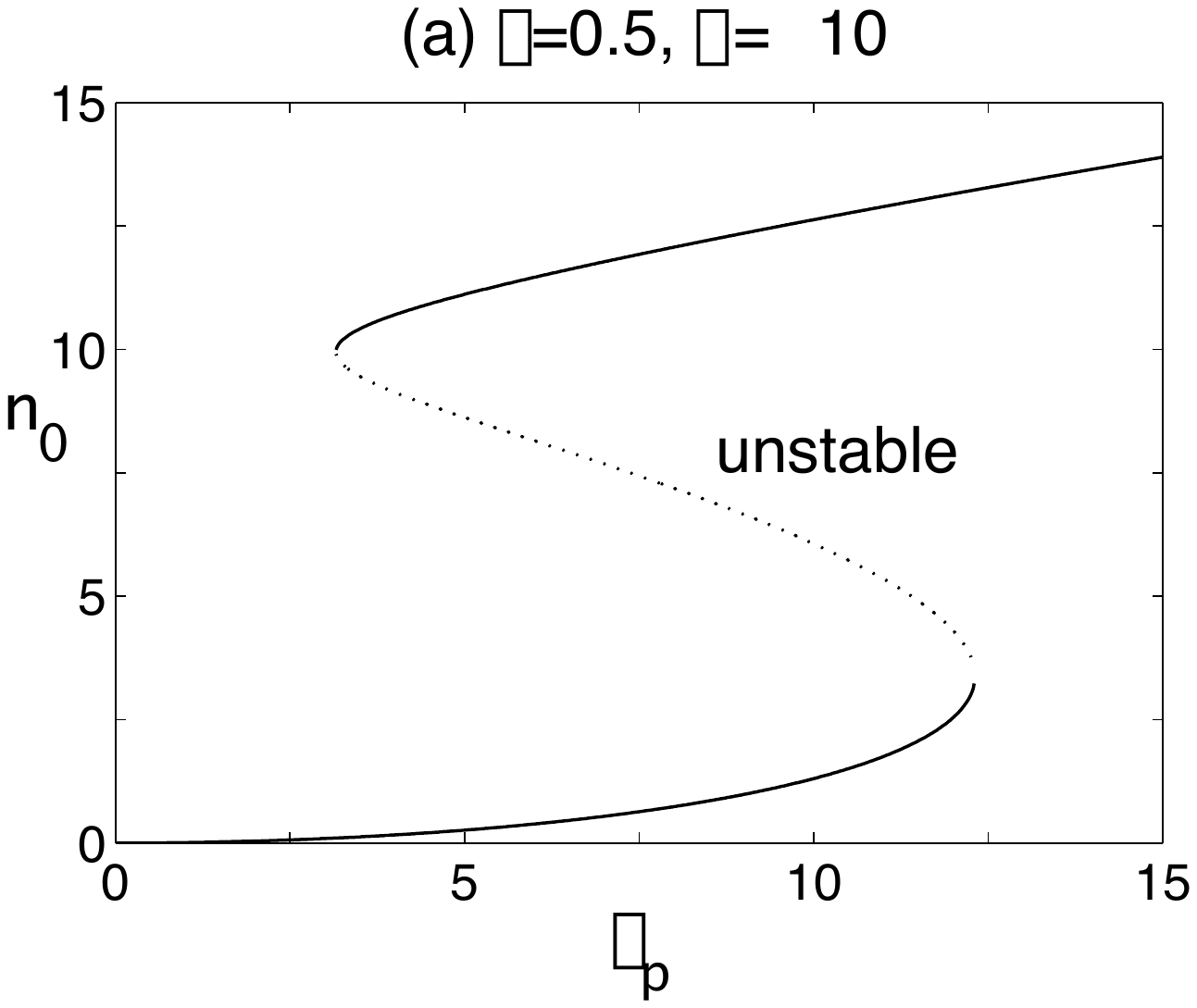}
\includegraphics[scale=0.5]{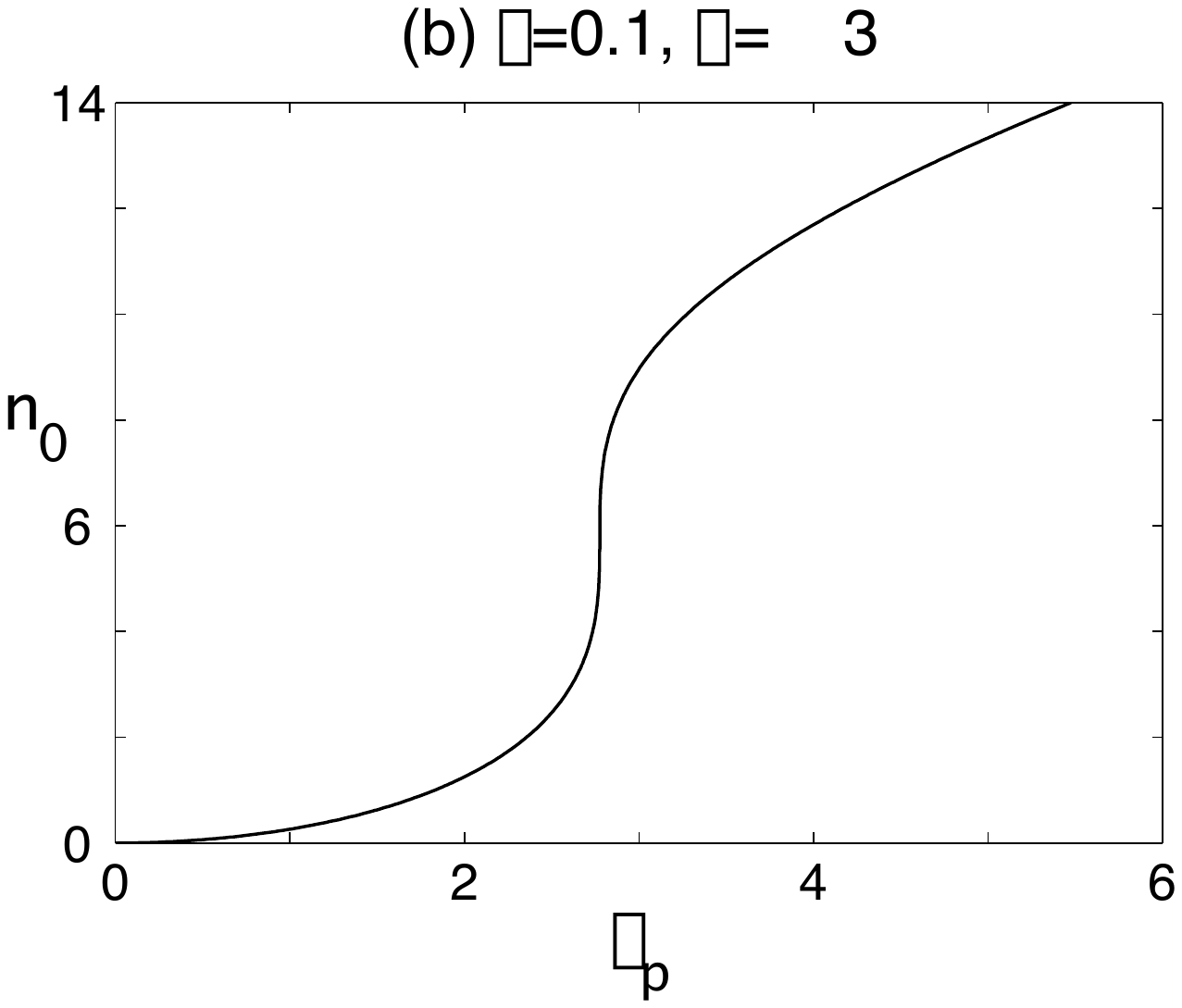}
\caption{Plots of the mean vibrational excitation number of the nanomechanical resonator, $n_0$ versus pump field intensity, $\epsilon_p$ for $\gamma=2.0$. The unstable branch shown in (a) is absent from (b) due to different values of pump field detuning and dispersion.} 
\label{fig1}
\end{figure}

In Figure \ref{fig1} we plot $n_0$ versus the pump intensity $\epsilon_p$ for various values of $\Delta$. Clearly under some conditions $n_0$ becomes a multi valued function of $\epsilon_p$.  In fact it can be shown that this will occur for negative detuning, $\Delta <0$. 
Not all the fixed point solutions are stable. To determine stability we linearise the equations of motion around the fixed points by writing $\alpha(t)=\alpha_0+\delta\alpha(t)$. The equations of motion for the fluctuation field $\delta\alpha(t)$ are then given by
\begin{equation}
\frac{d}{dt}\left (\begin{array}{c}
						\delta \alpha\\
						\delta\alpha^*
						\end{array}
						\right ) = M\left (\begin{array}{c}
						\delta \alpha\\
						\delta\alpha^*
						\end{array}
						\right )
\end{equation}
where 
\begin{equation}
M = \left (\begin{array}{cc}
						-\frac{\gamma}{2}-i(\Delta+4\chi n_0) & -iG\\
						iG^* & -\frac{\gamma}{2}+i(\Delta+4\chi n_0)
						\end{array}
						\right )
\end{equation}
where $G=2\chi \alpha_0^2$ and $\alpha_0$ is the solution to
\begin{equation}
\alpha_0\left [\frac{\gamma}{2}+i(\Delta+2\chi n_0) \right]=-i\epsilon_p
\end{equation}
Then we can write $\alpha_0=\sqrt{n_0}e^{i\phi_0}$ where
\begin{equation}
\tan \phi_0 = \frac{\gamma}{2\Delta+4\chi n_0}
\label{steadystate-phase}
\end{equation} 
As we have taken $\epsilon_p$ as real this is the phase shift of the oscillator field from the pump field. 

The eigenvalues of the linearised motion determine stability. These are given by
\begin{equation}
\lambda^{\pm}=-\frac{\gamma}{2}\pm i\sqrt{(\Delta+6\chi n_0)(\Delta+2\chi n_0)}
\end{equation}
For stability the real parts of these eigenvalues must be negative. The fixed points are unstable between the turning points of the state equation, Eq. (\ref{state-eqn}).    
 In Figure \ref{fig1} we show the unstable fixed points as a dashed line in plot (a). Note that from Eq.(\ref{turning}), 
 \begin{equation}
 |\lambda^{\pm}|^2\equiv\lambda^2=\frac{dI_p}{dn_0}
 \end{equation}
 and the eigenvalues vanish at the turning points. The linearised analysis thus breaks down at the bifurcation points (switching points)  

\section{Amplifier Gain}
Let us now turn to the operation of this device as an amplifier by including the weak signal field, $\epsilon_s\neq 0$. The pump field is chosen to pick out a particular steady state operating condition. We then include the signal field in the linearised equations of motion around that steady state. In an interaction picture, the equations of motion then become
\begin{equation}
\label{signal-eqn}
\frac{d}{dt}\left (\begin{array}{c}
						\delta \alpha\\
						\delta\alpha^*
						\end{array}
						\right ) = M\left (\begin{array}{c}
						\delta \alpha\\
						\delta\alpha^*
						\end{array}\right )
						+\left (\begin{array}{c}
						-i\epsilon_s e^{-i\delta t}\\
						i\epsilon_s e^{i\delta t}
						\end{array}\right )
\end{equation}
 where $\delta=\omega_s-\omega_p$ is the detuning between the signal frequency and the pump.  These linear equations are easily solved. Here we will be primarily be interested in the signal to noise ratio as a function of frequency in the long time limit. To this end we take the Fourier transform of  Eq. (\ref{signal-eqn}) and neglect initial conditions (which decay to zero in the long time limit). Define
 \begin{equation}
\delta  \tilde{\alpha}(\omega)=\int_{-\infty}^\infty e^{i\omega t} \delta \alpha(t)
 \end{equation}
 The equations of motion then imply that
 \begin{equation}
 \left (\begin{array}{c}
						\delta  \tilde{\alpha}(\omega)\\
						\delta  \tilde{\alpha}^*(\omega)
						\end{array}\right ) ={\cal G}(\omega) \left (\begin{array}{c}
						i\epsilon_s \delta(\omega-\delta)\\
						-i\epsilon_s \delta(\omega+\delta)
						\end{array}\right )
\label{gain}
\end{equation}
 where the gain metrix ${\cal G}(\omega)$ is defined by
 \begin{equation}
 {\cal G}(\omega)=(M+i\omega I)^{-1}
 \end{equation}
 with $I$ the identity matrix in two dimensions. Then
 \begin{equation}
 {\cal G}(\omega)=\left [4(\tilde{\Delta}^2-|G|^2)+(\gamma-2i\omega)^2\right ]^{-1}\left (\begin{array}{cc}
 										-2\gamma+4i(\omega+\tilde{\Delta}) & 4iG\\
										-4iG^* & -2\gamma+4i(\omega-\tilde{\Delta})\end{array}\right )
\label{gain_matrix}
\end{equation}
 $\tilde{\Delta}=\Delta+4\chi n_0$. Note that for $\chi=0$, ${\cal G}$ is diagonal corresponding to a driven and damped linear oscillator. 
 
The delta function frequency dependance on the right hand side of Eq. (\ref{gain}) is of course a consequence of the assumed monochromatic harmonic time dependance of the signal field. In general we would have a spectrum for the signal $\tilde{\epsilon}(\omega)$ in place of $\epsilon\delta(\omega+\delta)$.  In the case considered here the signal is harmonic and oscillating at frequency $\pm\delta$ (in the interaction  picture). Explicitly, the long time solution in the time domain is
\begin{equation}
\delta\alpha(t)= i\epsilon_s{\cal G}_{11}(\delta) e^{-i\delta t}-i\epsilon_s {\cal G}_{21}^*(\delta)e^{i\delta t}
\end{equation}

\begin{figure}[h]
\includegraphics[width=5cm]{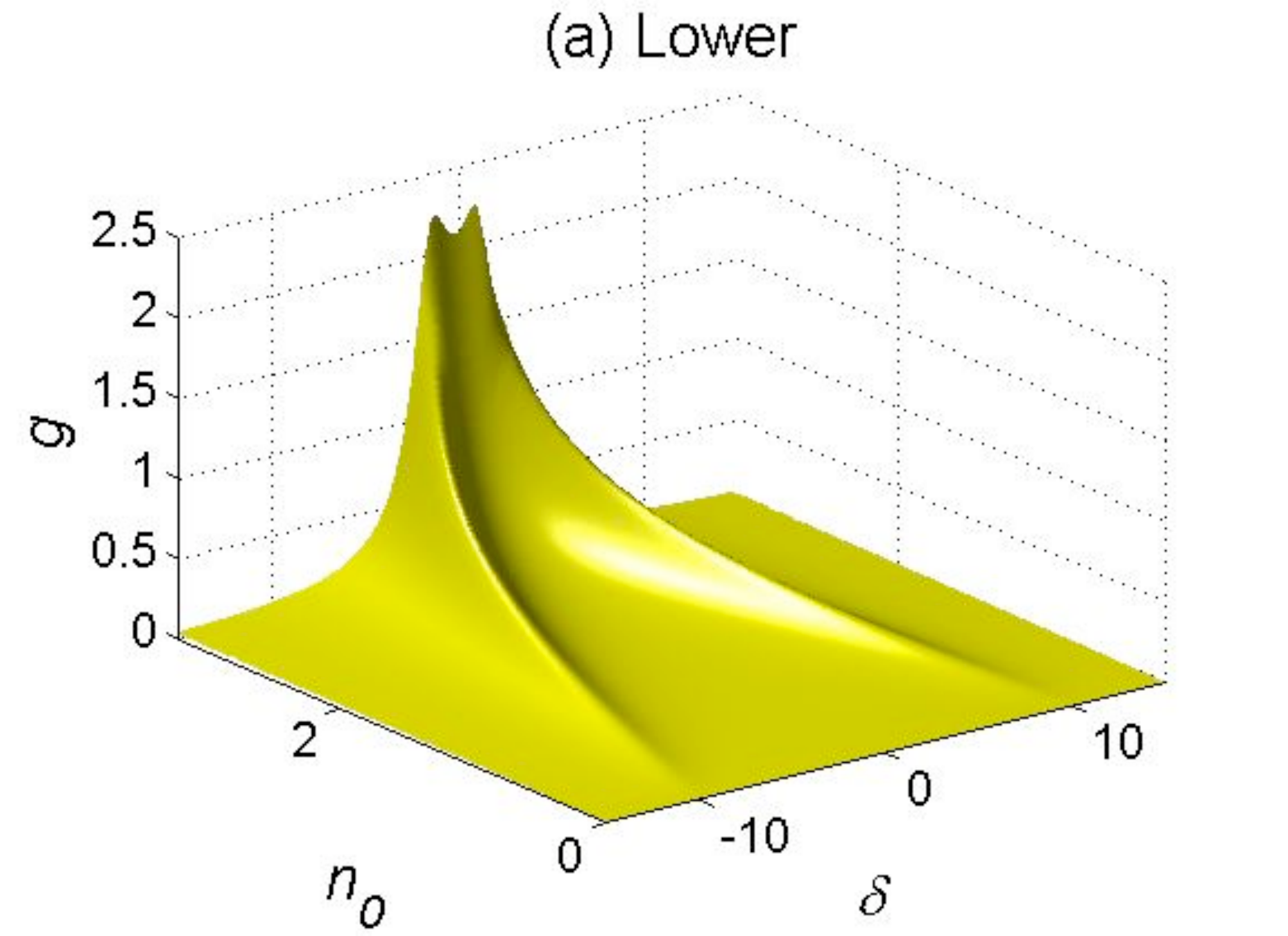}
\includegraphics[width=5cm]{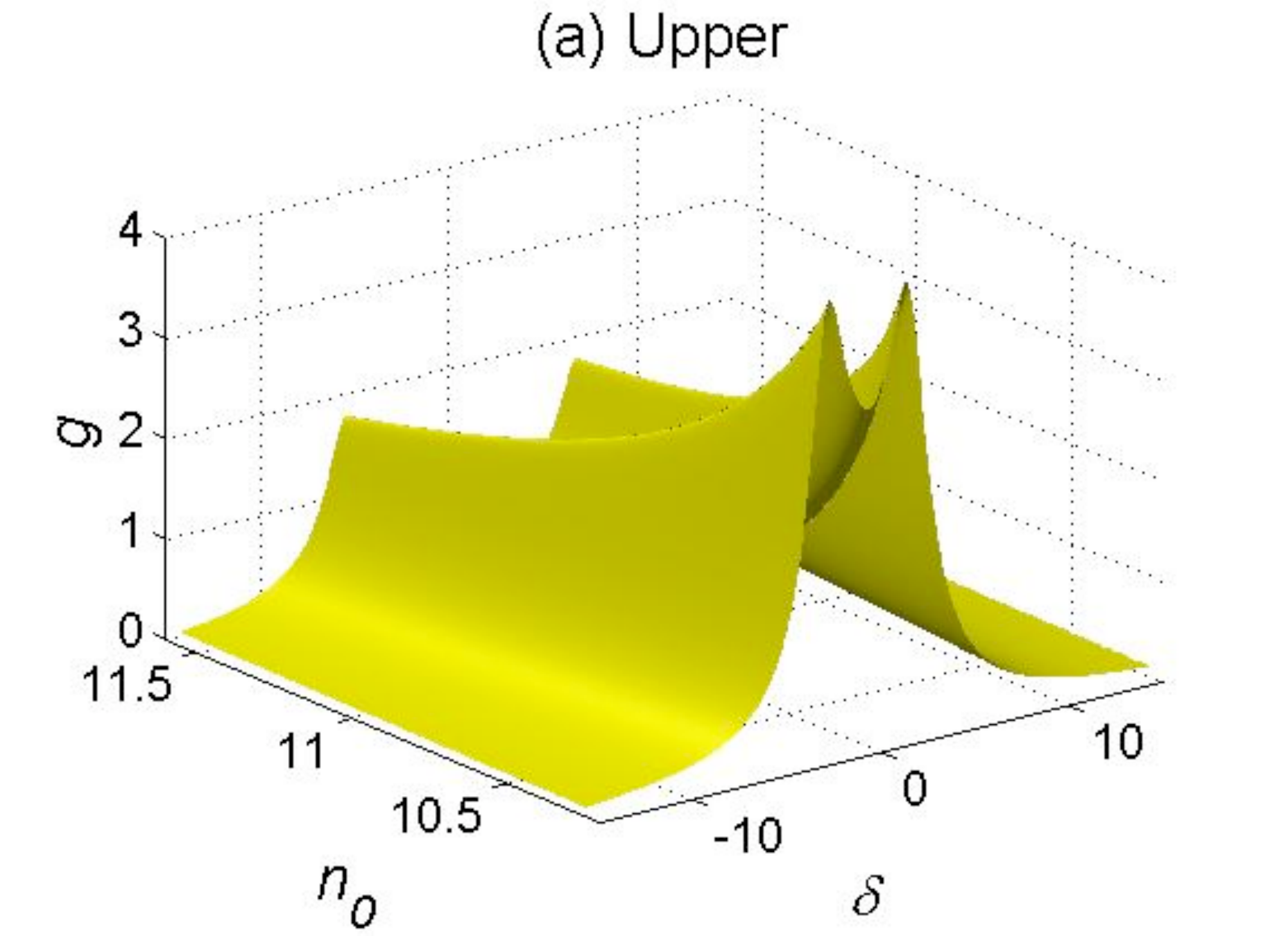}\\
\includegraphics[width=5cm]{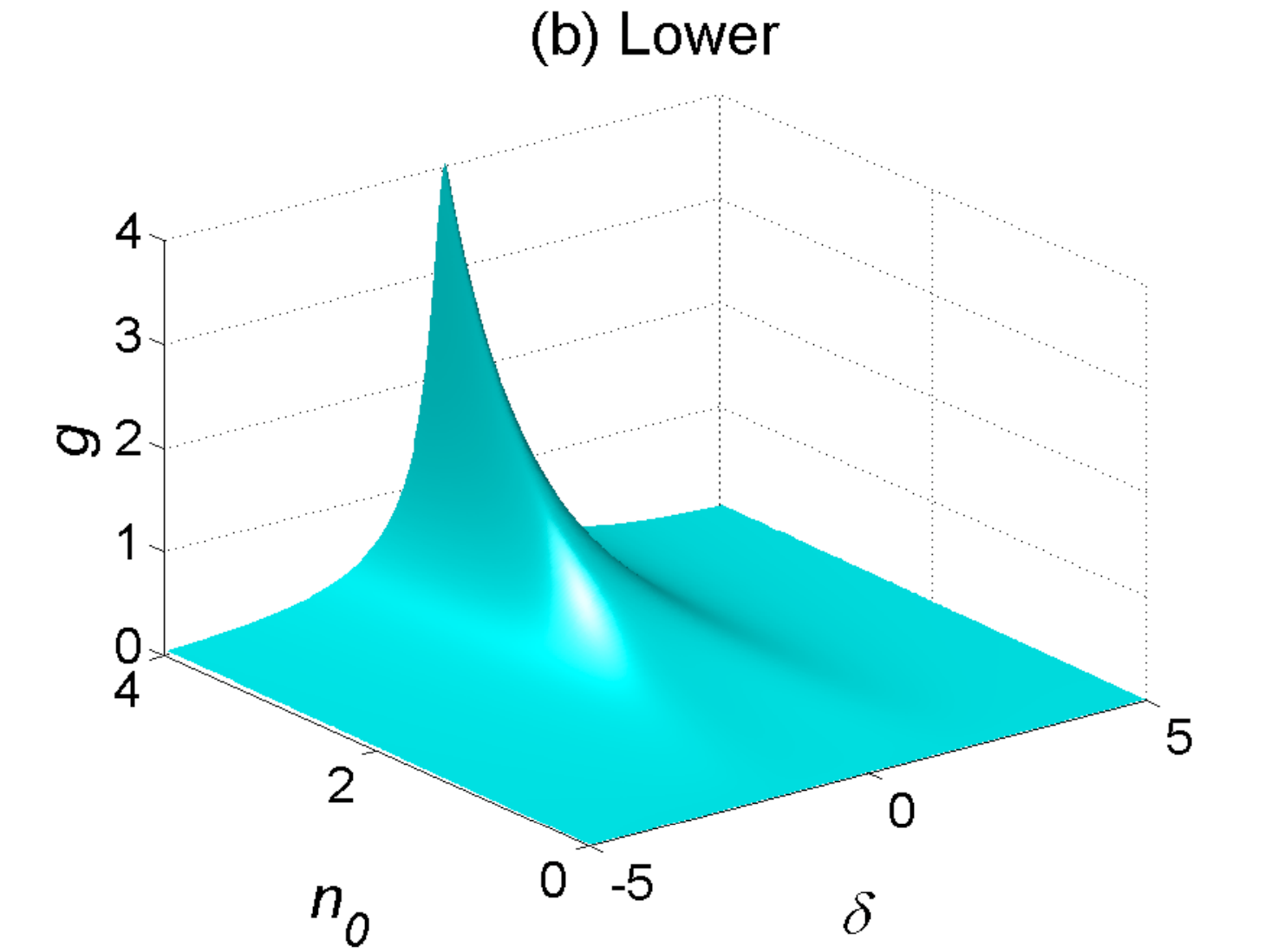}
\includegraphics[width=5cm]{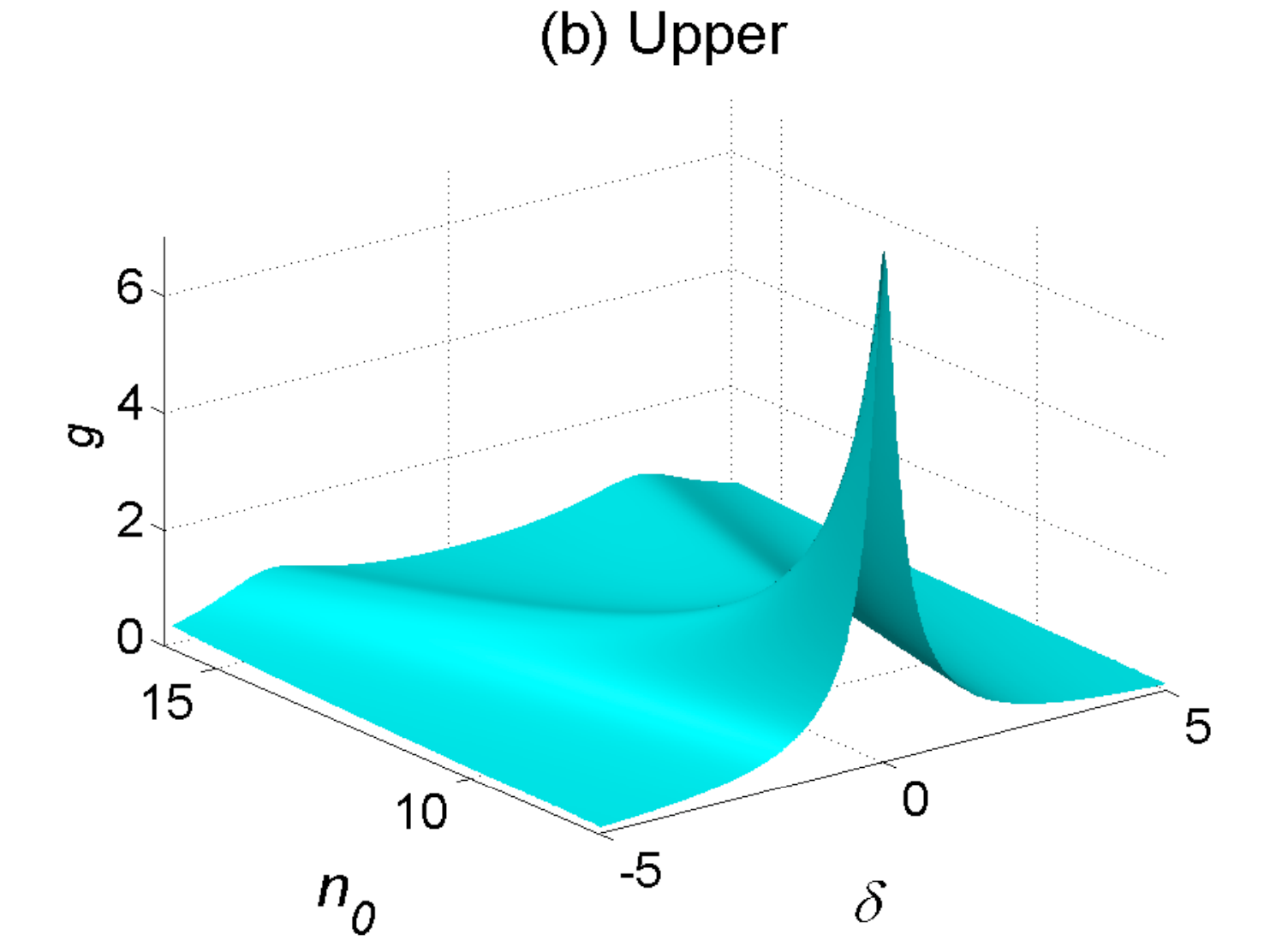}
\caption{Plots of $g(\delta,n_0)$ versus detuning and $n_0$, for the response curves in figure 1 (a) and (b).  We have separated out the the results for the lower and upper branches in order to avoid the region around the critical points where the linearisation breaks down. } 
\label{fig2}
\end{figure}

The appropriate signal quantity is the average of the linearised variable $\delta \hat{X}_\theta(t)\equiv \hat{X}_\theta(t)-\langle X_\theta\rangle_{0}$. To be specific we take the local oscillator frequency to be $\omega_{LO}=\omega_p-\delta$. The average signal then is most conveniently represented in terms of the Fourier amplitude of the mean, 
 \begin{eqnarray}
\delta \tilde{x}_\theta(\omega) & = & \int_{-\infty}^\infty dt e^{i\omega t} \langle \delta \hat{X}_\theta(t)\rangle\\
    & = & \delta\tilde{\alpha}(\omega-\delta)e^{i\theta}+ \delta\tilde{\alpha}^*(\omega+\delta)e^{-i\theta} 
    \end{eqnarray}
where $\delta\tilde{\alpha}(\omega)$ is the Fourier transform of $\delta\alpha(t)=\langle  \delta a(t)\rangle$ with $\delta a(t)= a_I(t)-\alpha_0$.  The homodyne signal is at DC in this picture so we evaluate this at $\omega=0$, we find that
\begin{equation}
  \delta \tilde{x}_\theta(0)  =  -i\epsilon_s\left \{[{\cal G}(\delta)]_{12} e^{i\theta}-[{\cal G}(-\delta)]_{21}e^{-i\theta}\right \}
\end{equation}
 where $[{\cal G}(\omega)]_{ij}$ is a matrix element of ${\cal G}(\omega)$. Using Eq. (\ref{gain_matrix}) this becomes
 \begin{equation}
   \delta \tilde{x}_\theta(0)= \frac{8\epsilon_s\chi\alpha_0^2 e^{i\theta}}{4(\tilde{\Delta}^2-|G|^2)+(\gamma-2i\delta)^2}+\mbox{c.c.}
   \end{equation}
   where c.c stands for complex conjugate. If we write $\alpha_0^2=n_0e^{2i\phi_0}$ this becomes
   \begin{equation}
     \delta \tilde{x}_\theta(0)= \frac{4\epsilon_s\chi n_0\left [(\lambda^2-\delta^2)\cos(\theta+2\phi_0)-\gamma\delta\sin(\theta+2\phi_0)\right ]}{(\lambda^2-\delta^2)^2+\gamma^2\delta^2}
      \label{nonlinear-quad}
      \end{equation}
         where 
   \begin{equation}
   \lambda^2=\frac{\gamma^2}{4}+(\Delta+6\chi n_0)(\Delta+2\chi n_0)
   \end{equation}
   Note that this quadrature amplitude goes to zero when the nonlinearity is turned off. We can thus say that this field component is generated by the nonlinear response of the cavity. 
If we now define 
 \begin{equation}
 \tan\nu=-\frac{\gamma\delta}{\lambda^2-\delta^2}
 \label{gain-angle}
 \end{equation}
 we see that 
    \begin{equation}
     \delta \tilde{x}_\theta(0)=\frac{4\epsilon_s\chi n_0}{\sqrt{(\lambda^2-\delta^2)^2+\gamma^2\delta^2}}\cos(\nu-2\phi_0-\theta)
     \end{equation}
     Thus for the appropriate choice of local oscillator phase at each bias point, $\theta=\nu-2\phi_0$,  the maximum gain can be defined as 
     \begin{equation}
 g(\delta,n_0)=    \frac{4\epsilon_s\chi n_0}{\sqrt{(\lambda^2-\delta^2)^2+\gamma^2\delta^2}}
 \end{equation} 
   This function is peaked along the lines $\lambda=\pm \delta$. In Figure \ref{fig2} we plot the gain versus the detuning and $n_0$ for this optimal choice of phase and with $\epsilon_s=\gamma=1$.   Note that for $\delta=0$ the gain is inversely proportional to square of the eigenvalues of the linearised dynamics. As these eigenvalues approach zero at the critical points the gain diverges at this point signaling a break down of the linear approximation. A full nonlinear analysis would be required to correctly compute the response near the switching points.   
Determining the conditions for maximum gain is not sufficient to specify the operating conditions as the noise might also be expected to peak around the bifurcation points. In the next section we turn to a linearised noise analysis to determine the signal to noise ratio for the model.  

For completeness we now give the signal when the local oscillator is tuned to $\omega_{LO}=\omega_p+\delta$. Proceeding as above we find the DC component of the average signal is now given by 
\begin{equation}
\delta \tilde{x}_\theta(0)=i\epsilon_s({\cal G}_{11}(\delta)e^{i\theta}-{\cal G}_{22}(-\delta)e^{-i\theta})
\end{equation}
This quadrature does not depend on the mean field in the cavity, $\alpha_0$. It thus  carries the empty cavity response. In fact if we set the pump to zero, turning of the nonlinearity, this quadrature is given by
\begin{equation}
\delta \tilde{x}_\theta(0)\rightarrow \frac{2\epsilon_s}{\sqrt{\frac{\gamma^2}{4}+\delta^2}}
\label{empty-quad}
\end{equation}
which is the empty cavity response to the driving field.

 \section{Noise Analysis}
 In order to compute the stationary noise power spectrum for the measured signal we again face the problem of not having a specific model of the transducer. However we can proceed in general terms by assuming that the transducer is described by a bosonic field and the nanomechanical resonator couples the input to this field,  $a_{in}(t)$ to the output $a_{out}(t)$ as discussed in the introduction. At high frequencies and for weak coupling the relation between the transducer fields and the amplitude of the mechanical resonator is given by Eq.(\ref{input-output}). 

The linearised quantum stochastic differential equations are then 
\begin{equation}
\label{qsde}
\frac{d}{dt}\left (\begin{array}{c}
						\delta a\\
						\delta a^\dagger
						\end{array}
						\right ) = M\left (\begin{array}{c}
						\delta a\\
						\delta a^\dagger
						\end{array}\right )
						+\sqrt{\gamma}\left (\begin{array}{c}
						a_{in}(t)\\
						a_{in}^\dagger(t)
						\end{array}\right )
\end{equation}
where we assume that the input transducer field is in a coherent state with vacuum noise only,
\begin{eqnarray} 
\langle a_{in}\rangle & = & (\epsilon_s/\sqrt{\gamma})e^{-i\delta t}\\
\langle a_{in}(t), a_{in}^\dagger(t')\rangle & = & \delta(t-t')
\end{eqnarray}
The complex amplitude of the signal field is coherent so it does not contribute to the noise. In what follows we assume that this amplitude has been subtracted off the detected transducer signal for the purposes of computing the noise. We thus drop it in the rest of this calculation and take the input transducer field to be in the vacuum state. 
 
  Drawing again on the quantum optical analogy with homodyne detection we will simply define a  stationary noise power spectrum as\cite{GardZol}
 \begin{equation}
 S^{out}_{\theta}(\omega)\equiv \int_{-\infty}^\infty dt e^{i\omega t} \langle\ :\ \delta \hat{X}^{out}_\theta(t),\delta\hat{X}^{out}_\theta(0)\ :\ \rangle 
 \label{noise-spectrum}
 \end{equation}
 where the colons denote normal ordering. In effect this means that a shot noise floor has been subtracted off the noise power spectrum. In terms of the Fourier component operators thus may be written as 
  \begin{equation}
 S^{out}_{\theta}(\omega)\equiv \int_{-\infty}^\infty d\omega' \langle\ :\ \delta \hat{X}^{out}_\theta(\omega),\delta\hat{X}^{out}_\theta(\omega')\ :\ \rangle 
 \end{equation}
 where 
 \begin{equation}
 \delta\hat{X}^{out}_\theta(\omega)= \delta \tilde{a}_{out}(\omega-\delta)e^{i\theta }+\delta \tilde{a}_{out}^\dagger(\omega+\delta)e^{-i\theta }
 \end{equation}
 Note that when $\epsilon_s=0, \delta=0$, Eq.(\ref{noise-spectrum}) is simply the  quadrature squeezing spectrum computed in \cite{Collett_Walls}. However here we are interested in the noise at the signal frequency, i.e. $\omega_s$.
 
 We can relate the output quadrature operators of the transducer to the quadrature operators of the nanomechanical resonator by taking the Fourier transform of Eq. (\ref{qsde}). Combining this with the relation in Eq.(\ref{input-output}) we can compute the noise power spectrum of the detected signal at the signal carrier frequency $\delta$. This corresponds to evaluating the output spectrum at $\omega=0$.  
  
Expanding the moments in terms of the Fourier amplitudes of the field fluctuation operators,  and applying the appropriate commutation relations, the spectrum can be expressed in terms of the gain matrix elements,
  \begin{eqnarray}
 S^{out}_{\theta}(\omega,\delta) & = & {\gamma}^2(
	e^{2i\theta}({\cal G}_{11}(\omega -\delta)+1/\gamma){\cal G}_{12}(-\omega +\delta)   
	+ {\cal G}_{21}(-\omega +\delta){\cal G}_{12}(\omega -\delta\nonumber)\\
	&&+ {\cal G}_{21}(\omega +\delta){\cal G}_{12}(-\omega -\delta)
	+ e^{-2i\theta}{\cal G}_{21}(\omega +\delta)({\cal G}_{22}(-\omega -\delta) +1/\gamma)  )
 \label{spectrum}
 \end{eqnarray}
Evaluating at $\omega=0$ gives the noise power in the homodyne signal as a function of signal detuning, 
\begin{equation}
 S^{out}_{\theta}(0,\delta)  =  \frac{ 2\gamma^2 |G|^2+\gamma\left (
	-iGe^{2i\theta}((\frac{\gamma}{2}-i\tilde{\Delta})^2+\delta ^2+|G|^2)+ c.c\right )}{(\lambda^2-\delta^2)^2+\gamma^2\delta^2}
 \end{equation}
We can recover the known results for the squeezing in this model\cite{Collett_Walls} when there is no injected signal by setting $\delta=0$ in Eq. (\ref{spectrum}) and setting the phase of the local oscillator, $\theta$ so that the on-resonance noise ($\omega=0$) is a minimum.

\begin{figure}[h]
\includegraphics[width=5cm]{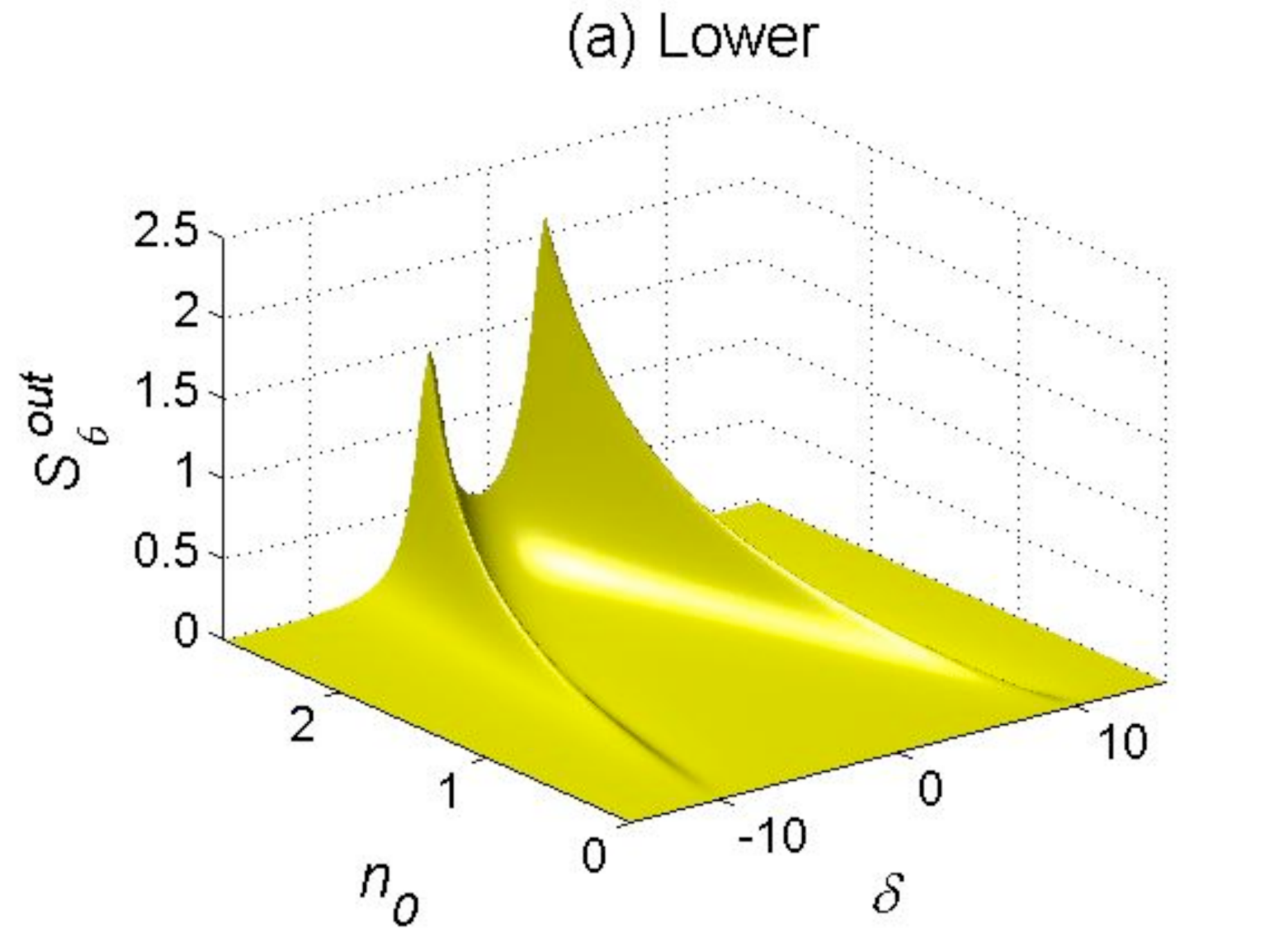}
\includegraphics[width=5cm]{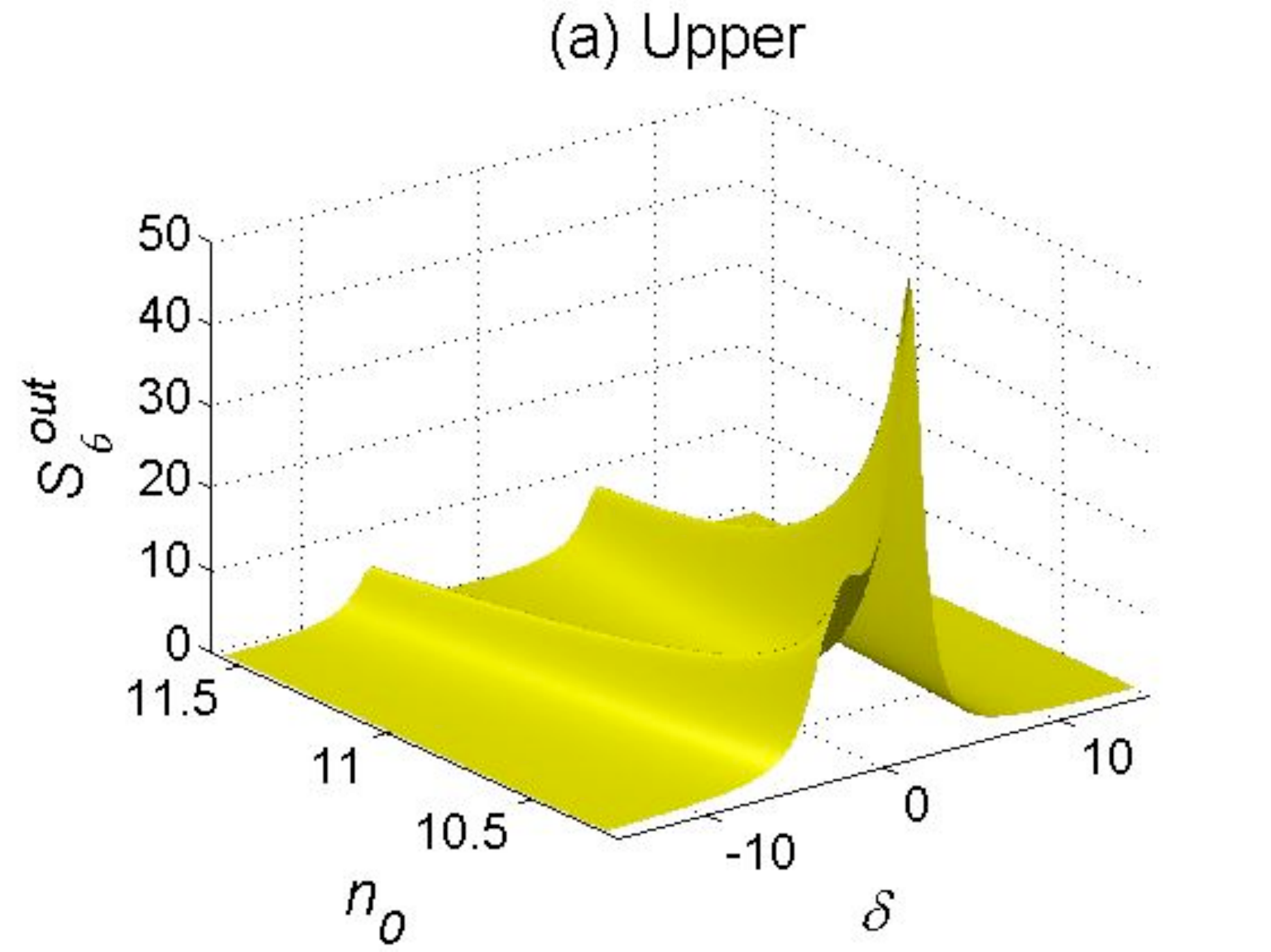}\\
\includegraphics[width=5cm]{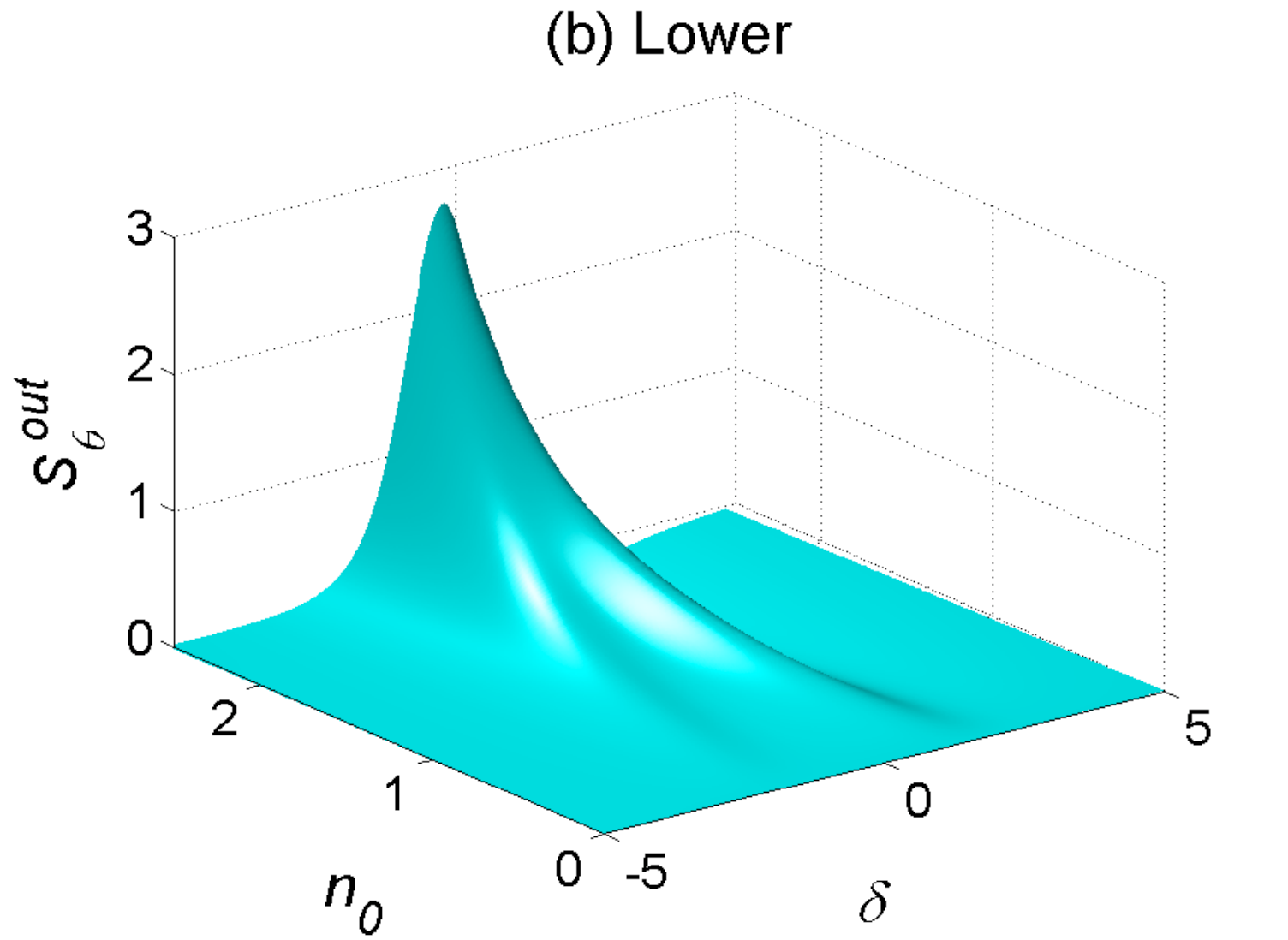}
\includegraphics[width=5cm]{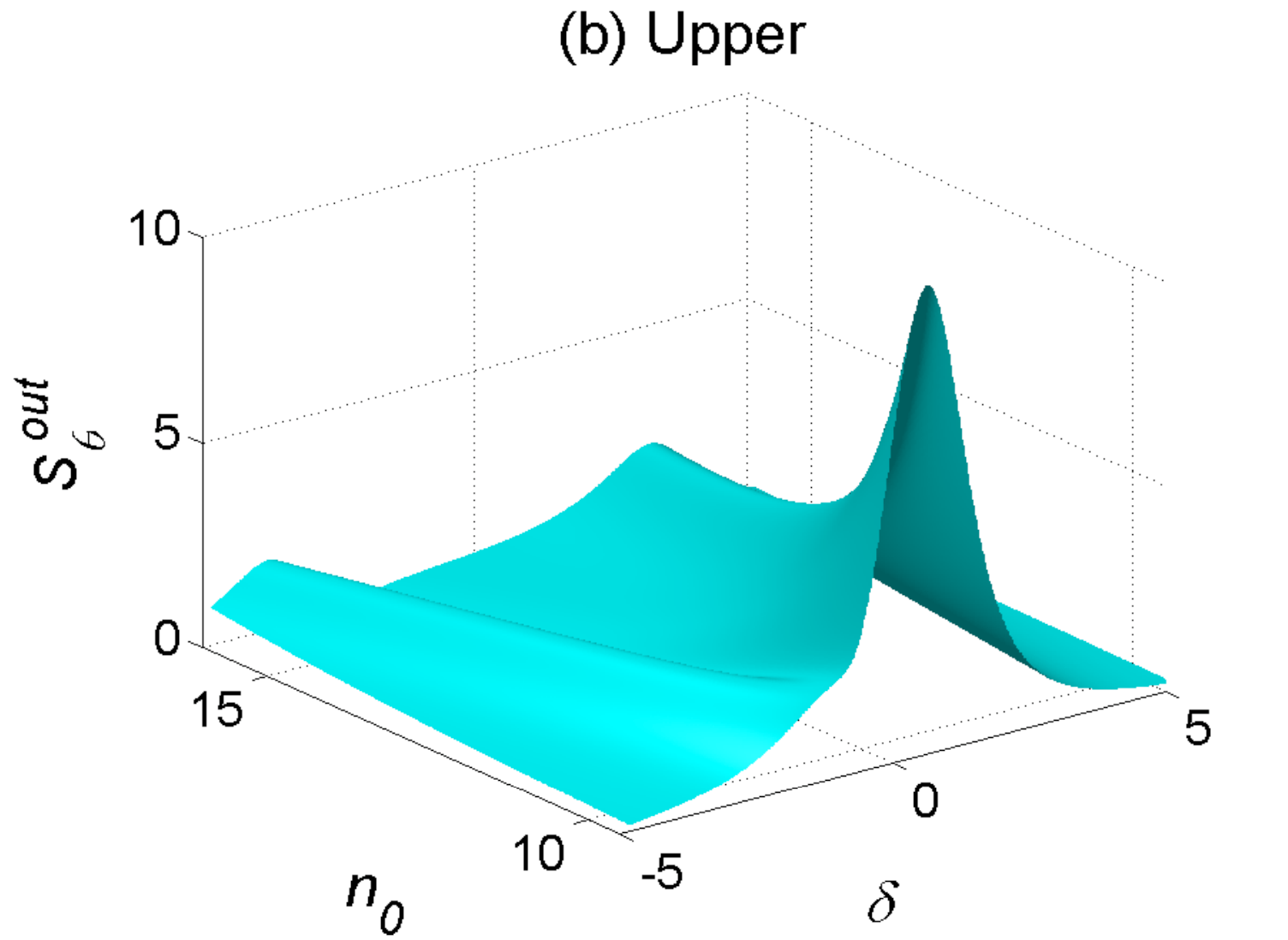}
\caption{Plots of the noise spectrum versus detuning between signal and pump fields, and $n_0$,  for the bias condition shown in  curve Figure 1 (a) and (b).  We have separated out the the results for the lower and upper branches in order to avoid the region around the critical points where the linearisation breaks down. } 
\label{fig3}
\end{figure}

Here we choose the particular local oscillator phase that was used to fix the gain of the average signal, $\theta=\nu-2\phi_0$, then
\begin{equation}
S^{out}_{\theta}(0,\delta) =  \frac{\gamma\left [8\chi^2 n_0^2\gamma+4\chi n_0(\delta^2-\lambda^2+\gamma^2/2)\sin(\nu-\phi_0)-4\chi n_0\gamma(\Delta+4\chi n_0)\cos(\nu-\phi_0)\right ]}{(\lambda^2-\delta^2)^2+\gamma^2\delta^2}
\end{equation}
The phase angles are given by Eq.(\ref{steadystate-phase}) and Eq.(\ref{gain-angle}). The output noise power spectrum as a function of the signal frequency is plotted, in Figure \ref{fig3}. In practice, other phase choices for the local oscillator might be preferred; for example one might choose the phase to find the quadrature with minimum noise. 


We can find an expression for the minimum detectable force. First define the square of the square of the  signal-to-noise-ratio (SNR), defined as
\begin{equation}
 SNR= \frac{|\langle \delta\hat{X}^{out}_\theta(\delta)\rangle|^2}{S_{\theta}(\delta)+1}
\label{snr}
 \end{equation}
Note that in this definition we have added the vacuum noise level to the normally ordered variance of the output variable to get the total noise level. It is clear that the gain is peaked on $\delta=\pm\lambda$ so we set this condition and find the minimum value of $\epsilon_s$ for which the SNR is unity. 

To evaluate performance we can compare the response of the signal to the nonlinearity in the cavity to the response when the nonlinearity is turned off, i.e. the cavity is empty. In this case, for our choice of signal phase, the response is given by the quadrature phase in Eq. (\ref{empty-quad})  found at dc for a local oscillator tuned to $\omega_p+\delta$. The noise in this quantity is simply vacuum noise, so the minimum detectable force, given by setting the signal to noise ratio to unity is,
\begin{equation}
\epsilon_s^{min}=\frac{1}{2}\left (\frac{\gamma^2}{4}+\delta^2\right )^{1/2}\ \ \ \ \mbox{(empty cavity)}
\end{equation}
This is clearly a minimum on resonance as expected, so that the minimum detectable force, in units of cavity line width is $\epsilon_s^{min}/\gamma=0.25$. 

\begin{figure}
\includegraphics[width=8cm]{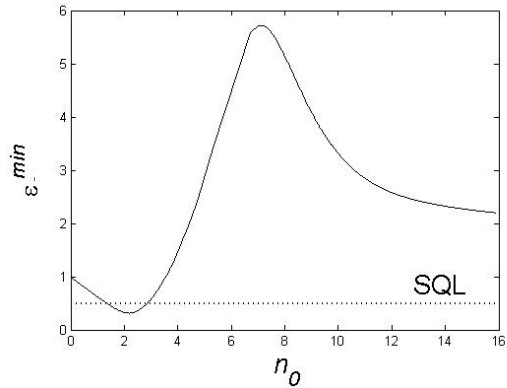}
\caption{A plot of the minimum detectable force versus $n_0$ for the mono stability case. The dashed line represents the minimum detectable force for an empty cavity, a standard quantum limit for this model.  } 
\label{minimum-detect}
\end{figure}

With the nonlinearity turned on, we consider the quadrature phase signal generated by the nonlinearity itself, Eq.  (\ref{nonlinear-quad}). There are many parameters that one might want to optimise for maximum signal to noise ratio. This is made difficult by the fact that the phase of the mean amplitude and the phase of the squeezed quadrature are not necessarily the same in this model. However for the choice of local oscillator phase we have used here, we see that the gain is maximised along the curves $\delta=\lambda$. Furthermore we expect quantum noise reduction inherent in this model to be good close to the instability at $\lambda\rightarrow0$. In this limit the minimum detectable signal squared is given by
\begin{equation}
\left (\epsilon_s^{min}\right )^2=\frac{\gamma^2}{4} \left(
\frac{8\chi^2n_0^2+\lambda^2}{\frac{\gamma^2}{4}+\lambda^2+4\chi^2n_0^2}
+2\chi Re\left(-\frac{i\alpha_0^2}{(\frac{\gamma}{2}-i\tilde{\Delta})}
\right)\right)
\end{equation}
In Figure \ref{minimum-detect} we plot $\epsilon_s^{min}$ versus $n_0$. The dotted line shows the best that can be achieved with the nonlinearity turned off. We clearly see a region of operation, close to the switching point, in which the minimum detectable force is less than the empty cavity case. 

\section{Conclusion}
In this paper we have proposed a model for the quantum noise of a driven nonlinear nanomechanical resonator. We have computed the small signal gain, and signal to noise ratio, for a driven and damped nonlinear nanomechanical system at zero temperature. As expected the noise is greatest when the system is pumped near the switching point between the upper and lower fixed point. Considered as a function of the detuning between the pump frequency and the small signal frequency, we see that there are optimal detunings where the signal to noise ratio is large.  For an appropriate choice of local oscillator phase reference, the gain and signal to noise ratio are maximised when the detuning between the pump and signal field is equal to the magnitude of the eigenvalues of the linearised dynamics near the fixed points. We show that there is a region of operation in which the minimum detectable force is less than it would be for a linear cavity. 
\acknowledgments
GJM would like to thank the Australian Research Council for their support.

\end{document}